\documentstyle[12pt,epsf]{article}
\begin{document}
\renewcommand{\thesection}{\Roman{section}}
\bibliographystyle{unsrt}
\def\rf{\bibitem}
\newcommand{\nwc}{\newcommand}
\nwc{\be}{\begin{equation}}
\nwc{\ee}{\end{equation}}
\nwc{\bea}{\begin{eqnarray}}
\nwc{\eea}{\end{eqnarray}}
\nwc{\ba}{\begin{array}}
\nwc{\ea}{\end{array}}
\nwc{\rtr}{\rangle}
\nwc{\ltr}{\langle}
\nwc{\ket}[1]{|#1\rtr}
\nwc{\bra}[1]{\ltr#1|}
\nwc{\0}{\ket{0}}
\nwc{\bu}{\ket{\uparrow}}
\nwc{\bd}{\ket{\downarrow}}
\nwc{\2}{\ket{\uparrow \downarrow}}
\nwc{\scal}[2]{\bra{#1}#2\rtr}
\nwc{\dagg}{\dagger}
\nwc{\emp}{\emphasize}
\nwc{\lb}{\label}
\nwc{\pa}{\partial}
\nwc{\paf}[2]{\frac{\pa#1}{\pa#2}}
\nwc{\ra}{\rightarrow}
\nwc{\Tr}{\mbox{\rm Tr}}
\nwc{\real}{\mbox{\rm Re}}
\nwc{\im}{\mbox{\rm Im}}
\nwc{\bino}[2]{\mbox{$\left(\begin{array}{c}#1\\#2\end{array}\right)$}}
\def\a{\alpha}
\def\b{\beta}
\def\d{\delta}
\def\e{\epsilon}
\def\l{\lambda}
\def\m{\mu}
\def\n{\nu}
\def\o{\omega}
\def\f{\phi}
\def\p{\pi}
\def\t{\theta}
\def\D{\Delta}
\def\O{\Omega}
\def\s{\sigma}
\nwc{\eps}{\epsilon}
\nwc{\br}{\mbox{\bf{R}}}
\nwc{\bc}{\mbox{\bf{C}}}
\nwc{\cz}{{\cal Z}}
\nwc{\cd}{{\cal D}}
\nwc{\ch}{{\cal H}}
\nwc{\cu}{{\cal U}}
\nwc{\cs}{{\cal S}}
\nwc{\cl}{{\cal L}}
\nwc{\cq}{{\cal Q}}
\nwc{\ca}{{\cal A}}
\nwc{\nh}{{\widehat n}}
\nwc{\up}{\uparrow}
\nwc{\down}{\downarrow}
\nwc{\cdag}{c^{\dagg}}
\nwc{\bdag}{b^{\dagg}}
\nwc{\de}{\delta}
\nwc{\unit}{\hat{1}}
%%%%
%
% Math Symbols
%
%%%%
\def\inbar{\,\vrule height1.5ex width.4pt depth0pt}
\def\IG{\relax\,\hbox{$\inbar\kern-.3em{\rm G}$}}
\def\IU{\relax\,\hbox{$\inbar\kern-.3em{\rm U}$}}
\def\ID{\relax{\rm I\kern-.18em D}}
\def\IF{\relax{\rm I\kern-.18em F}}
\def\IH{\relax{\rm I\kern-.18em H}}
\def\II{\relax{\rm I\kern-.17em I}}
\def\I1{\relax{\rm 1\kern-.28em l}}
\def\IM{\relax{\rm I\kern-.18em M}}
\def\IN{\relax{\rm I\kern-.18em N}}
\def\IP{\relax{\rm I\kern-.18em P}}
\def\IQ{\relax\,\hbox{$\inbar\kern-.3em{\rm Q}$}}

\def\IC{\hbox{{\bf \inbar}\hskip-4.0pt C}}

\def\IR{\hbox{I\hskip-1.7pt R}}
\font\cmss=cmss10 \font\cmsss=cmss10 at 7pt
\def\IZ{\relax\ifmmode\mathchoice
{\hbox{\cmss Z\kern-.4em Z}}{\hbox{\cmss Z\kern-.4em Z}}
{\lower.9pt\hbox{\cmsss Z\kern-.4em Z}}
{\lower1.2pt\hbox{\cmsss Z\kern-.4em Z}}\else{\cmss Z\kern-.4emZ}\fi}
%
%%%
\def\~#1{{\mathaccent"7E #1}}
\thispagestyle{empty}
\setcounter{footnote}{0}

\hfill \today

\vskip2cm

\centerline {\Large {\bf PHASE SEPARATION AND THREE-SITE HOPPING  }}
\vspace{5mm}
\centerline {\Large {\bf IN THE 2-DIMENSIONAL t-J MODEL}}
\vspace{5mm}
\centerline{\Large  {\bf{}}}
\vspace{0.75cm}
\centerline {Elisa Ercolessi$^{a)}$,
             Pierbiagio Pieri$^{a,b)}$, 
             Marco Roncaglia$^{a,b)}$ }
\vspace{1cm}
\centerline {\it $^{a)}$Dipartimento di Fisica, Universit\`a di
Bologna and INFM }
\centerline {\it Via Irnerio 46, I-40126, Bologna, Italy.}
\centerline {\it $^{b)}$INFN, Sezione di Bologna, Italy.}

\vspace{.5cm}

\begin{abstract}
We study the $t$-$J$ model with the inclusion of the so called three-site 
term which comes out from the $t/U\to 0$ expansion of the Hubbard model.
We find that this singlet pair hopping term has no qualitative effect 
on the structure of the pure mean field phase diagram for nonmagnetic states. 
In accordance with experimental data on high-$T_c$ materials 
and some numerical studies, we also find wide regions of phase 
coexistence whenever the coupling $J$ is greater than a critical value $J_c$.  
We show that $J_c$ varies linearly with the temperature $T$, going to 
zero at $T=0$.


\end{abstract}
\newpage

\setcounter{footnote}{0}

In a recent paper \cite{EPR} we have presented a new finite
temperature phase diagram for the nonmagnetic phases (dimer
and flux) in the 2-dimensional $t$-$J$ model below half-filling, 
which shows the existence of broad areas of phase coexistance 
(dimer-flux or flux-uniform), in accordance with experimental and 
numerical data on the possibility of separation into hole-rich and 
hole-poor regions \cite{DR,EKL,gri}. 
 
It is well known that one can obtain the $t$-$J$ model in the strong
coupling regime ($U\gg t$) of the one-band Hubbard model:
\be \lb{1bh}
H = - t \sum_{(ij)}\sum_\sigma c^\dagg_{i\s} c_{j\s}+ U \sum_i
n_{i \uparrow} n_{i\downarrow } 
\ee
below half-filling. In (\ref{1bh}) Latin indices label sites, the $c_{i\sigma
}$'s are annihilation operators for a fermion on site $i$ and
spin $\sigma=\uparrow ,\downarrow $, $n_{i\s}=$ $c^\dagg_{i\s} c_{i\s}$ and
$(ij) $ denotes a sum over ordered nearest neighbor (n.n.) pairs. 

The strong coupling limit of (\ref{1bh}) can be studied either by direct
perturbation methods \cite{E} or by  canonical transformations
methods \cite{Aq} (see also \cite{Boo,Vie} for a detailed review). 
To lowest order in $t/U$, the Hamiltonian that describes the
Hubbard model on the subspace with no double occupancies ($n_i \leq 1$) is
given by:
\be
{\widehat H} = T_h + H^{(1)} + H^{(2)} \label{effective}
\ee
where
\bea
 T_h& =& - \sum_{(ij)\sigma}\, 
   t_{ij} \, (1-n_{i\bar\sigma})\, 
    c^\dagg_{i\sigma}\, c_{j\sigma}
     (1-n_{j\bar\sigma})\;\; ; \; 
      {\bar \sigma}=-\sigma               \label{eff-a}\\
 H^{(1)} &=&  \sum_{\langle ij\rangle}\, 
       J_{ij} \left\{
            \vec{S}_i \cdot \vec{S}_j - 
      \frac{1}{4}\, 
        n_{i} n_j
            \right\} \;\;  , \;    J_{ij} = 4 \frac{|t_{ij}|^2}{U} 
          \; ,     \label{eff-b} \\
 H^{(2)} &=& -\frac{1}{U} \sum_{(ijl) }\, 
            \sum_{\sigma\tau}\, t_{ij} \, t_{jl }\, 
              (1-n_{i\bar\sigma})\, 
               c^\dagg_{i\sigma}\, c_{j\sigma}\, 
      n_{j\bar\sigma}\, n_{j\bar\tau}\, 
               c^\dagg_{j\tau}\, c_{l\tau} 
       (1-n_{l\bar\tau})              \label{eff-c} 
\eea
and the symbol $\langle ij\rangle $ denotes a summation over unordered 
pairs of n.n. and $(ijl)$ denotes a summation where $i\not= l$, and both 
$i$ and $l$ are n.n. to  $j$. 
The spin operators are defined as
\be \lb{spin}
\vec{S}_i = \frac{1}{2} \sum_{\a\b} c^\dagg_{i\a} \vec{\s}_{\a\b}
c_{i\b} \; , 
\ee
with $\vec{\s}=(\s_1,\s_2,\s_3)$ being Pauli matrices.

The effective Hamiltonian contains a direct hopping term $T_h$ that 
gives nonvanishing contribution only if the electron jumps from site $i$ to an
empty n.n. site $j$. So we expect the effective hopping coefficient $t$ to be
reduced by a factor of order $\d$, $\d$ being the doping fraction \cite{and}.
On the contrary $H^{(1)}$, which gives rise to a spin dynamics described by 
the standard AFM Heisenberg model, has not to be renormalized, 
since $\vec{S}_i \cdot \vec{S}_j$ is nonzero only if both the sites $i$ and 
$j$ are singly occupied. 
$H^{(2)}$ is proportional to $|t|^2$ and describes a double hopping process.
Thus, one would be tempted to believe that it has to be renormalized by 
a factor of $\d^2$. 
However, it is not difficult to see that the three-site
hopping term $H^{(2)}$ gives nonzero contribution only for the situation in
which the final site $l$ is empty and the intermediate site $j$ is singly 
occupied. Then this term should be renormalized by a factor of $\d(1-\d)$. 
Similar conclusions can be reached within different approaches 
\cite{Spa,KL,GK}. 

For low doping the three-site term is smaller by a factor $\d$ as 
compared with the AFM term and by a factor $|t|/U$ as compared with the direct 
hopping term. 
That is why it is usually ignored and one is led to the 
$t$-$J$ model \cite{H}. However, the Hubbard Hamiltonian is often 
studied in a range 
of the parameter $t/U$ not so small, especially in applications to high T$_c$
materials for which one assumes $t/U\approx 0.1$. 
In this regime the role of the three-site term cannot be neglected a
priori.

Numerical works have recently shown that the the three-site term is 
essential to avoid discrepancies between the one-particle spectra of 
the Hubbard and $t$-$J$ model in the intermediate $U$ regime \cite{EE}.
According to \cite{EE}, only after the inclusion of the three-site term, 
the $t$-$J$ model can correctly describe the photoemission experiments 
on cuprate oxides.

The three-site term describes the hopping of a singlet pair, which can be 
either rigid or with a flip of the spins. Both these processes preserve 
the local singlet bond and may be relevant for superconducting correlations, 
favoring superconducting states at the expense of an antiferromagnetic 
phase \cite{Spa}.

Therefore, according to the discussion above, we approximate the effect of
the Gutzwiller projector by renormalizing each term of the Hamiltonian (2) 
with suitable factors depending on $\d$.  
Our new starting point is hence the Hamiltonian:

\be
\ch = t\d \sum_{(ij)} \sum_{\a} c^\dagger_{i\a} c_{j\a}
- \frac{J}{4} \sum_{ijkl}\;\sum_{\a\b} \;c^\dagger_{i\a} \;c_{j\a}\;
\ca_{ji,kl}\; c^\dagger_{l\b}\; c_{k\b} \label{tj}
\ee
where we have introduced the four fermion matrix

\be
\ca_{ji,kl} = \left\{  
\begin{array}{cl} 
1 & \mbox{if $i$ n.n. $j$ and $i=k$, $j=l$} \\
\displaystyle \d(1-\d) & \mbox{if $i$ n.n. $j$, $k$ n.n. $l$, 
$j=l$ and $i\not=k$} \\
0 & \mbox{otherwise} \end{array} \right. \; .    
\ee

In \cite{EPR}, we examined the so-called nonmagnetic phases for the $t$-$J$ 
model, first proposed by Affleck and Marston \cite{AM}, 
in the mean field approximation and we have presented a finite
temperature phase diagram ($T$ versus the doping $\d$). 
In this paper we shall study how the phase diagram changes if we include the 
three-site hopping term.

To study the Hamiltonian (\ref{tj}) we can parallel the discussion in 
\cite{EPR}, where we have chosen to represent the partition function
in the grand canonical ensemble by means of a functional integral over
Grassmann fields and decoupled the four-fermion interaction by means of a
Hubbard-Stratonovich transformations over the auxiliary fields $\cu_{ij}(\tau)
= \cu_{ji}^\dagger(\tau)$.  In the static approximation $\cu_{ij}(\tau)\equiv
\cu_{ij}$, the partition function reduces to:

\bea \label{pf} 
\cz &=& \int [\cd \cu^*_{ij} \cd \cu_{ij}]\, \exp\left\{ - \frac{\b}{J}
\sum_{ijkl}\;\cu^*_{ij} \;\ca_{ji,kl}\;\cu_{kl} \right\} \exp\left\{ -
\cs_{eff}\right\}  \\ 
\cs_{eff}&\equiv& - 2\, \mbox{Tr} \left\{ \log \left[ -
\IG^{-1} + \b \IU \right] \right\} \; ,
\eea
where 
\bea
[\IG^{-1}]_{nn'}^{ij} &= &( i\o_n + \m \b )\, \d_{ij} \d_{nn'} \\
\left[ \IU\right]_{nn'}^{ij} &=& \left\{  
\begin{array}{cl} 
[t\d + \displaystyle\sum_{pq}\;\cu_{pq}\;\ca_{pq,ji}] \d_{nn'} & 
\mbox{if $i$ n.n. $j$} \\ 
0 & \mbox{otherwise} \end{array} \right. \; ,
\eea
$\mu$ being the chemical potential.

We consider, as in \cite{AM}, 
a pattern of link variables $\cu_{ij}$ that admits a symmetry for
translations along the diagonal so that the matrix $\IU$ depends only on the
four parameters $\cu_j$ ($j=1,\cdots,4$) shown in Fig. 1 of \cite{EPR}. 
As for the pure $t$-$J$ model, the matrix $\IU$ can be easily diagonalized in
momentum space: if we include the three-site term it has eigenvalues

\bea E_k &=& \pm \,\, |\l_k| \label{eigen}\\ 
\l_k &\equiv& \chi_1
e^{ik_x a}+\chi_2^* e^{-ik_y a}+\chi_3 e^{-ik_x a}+ \chi_4^*  e^{ik_y a}
\nonumber 
\eea 
with the definitions 
\bea
\chi_1 &\equiv&  t\d + \cu_1 + \displaystyle\d(1-\d)
(\cu_2^*+\cu_3+\cu_4^*)\\
\chi_2 &\equiv&  t\d + \cu_2 + \displaystyle\d(1-\d)
(\cu_1^*+\cu_3^*+\cu_4)\\
\chi_3 &\equiv&  t\d + \cu_3 + \displaystyle\d(1-\d)
(\cu_1+\cu_2^*+\cu_4^*)\\
\chi_4 &\equiv&  t\d + \cu_4 + \displaystyle\d(1-\d)
(\cu_1^*+\cu_2+\cu_3^*)
\eea

We notice that both the expression for the partition function (\ref{pf}) 
and for the eigenvalues of $\IU$ (\ref{eigen}) are formally identical 
to the one for the pure $t$-$J$ model we gave in \cite{EPR}, 
the only differences being in the definition of the matrix $\ca$ and of the 
parameters $\chi_j$'s that appear in $\lambda_k$. 
Thus, to study nonmagnetic phases in the mean field approximation, 
we can borrow from the discussion of \cite{EPR}. 
%%%%%%%%% frase cambiata
%We have solved numerically the self-consistency equations 
%for the case $t/J=1$. 

We first discuss in detail the numerical solutions to the 
self-consistency equations for the case $J/t=1$.
For high temperature and/or high doping the free energy 
is minimized by the uniform phase, characterized by 
$\cu_i\equiv \cu\in \IR^{+}$, $i=1,\dots,4$. For temperature lower than $T=J/4$ 
we find two phases: dimer and flux. The former corresponds to two 
solutions of the form $\cu_1 \neq \cu_2 = \cu_3 = \cu_4 \; , \; \cu_j \in \IC$ 
with $|\cu_1| \gg |\cu_2|$ and is stable in the lower range of dopings. 
In the flux phase the link variables $\cu_i$ are all equal to a complex 
value $\cu e^{i\f}$ with $\f \neq 0$. 
It is the lowest energy state in a range of doping that roughly 
corresponds to that of high-$T_c$ superconductivity. 
The pure mean field diagram is shown in Fig. 1 (solid triangles).

\begin{figure}[htbp]
\begin{center}
\leavevmode
   \epsfxsize=0.35\textwidth\epsfbox{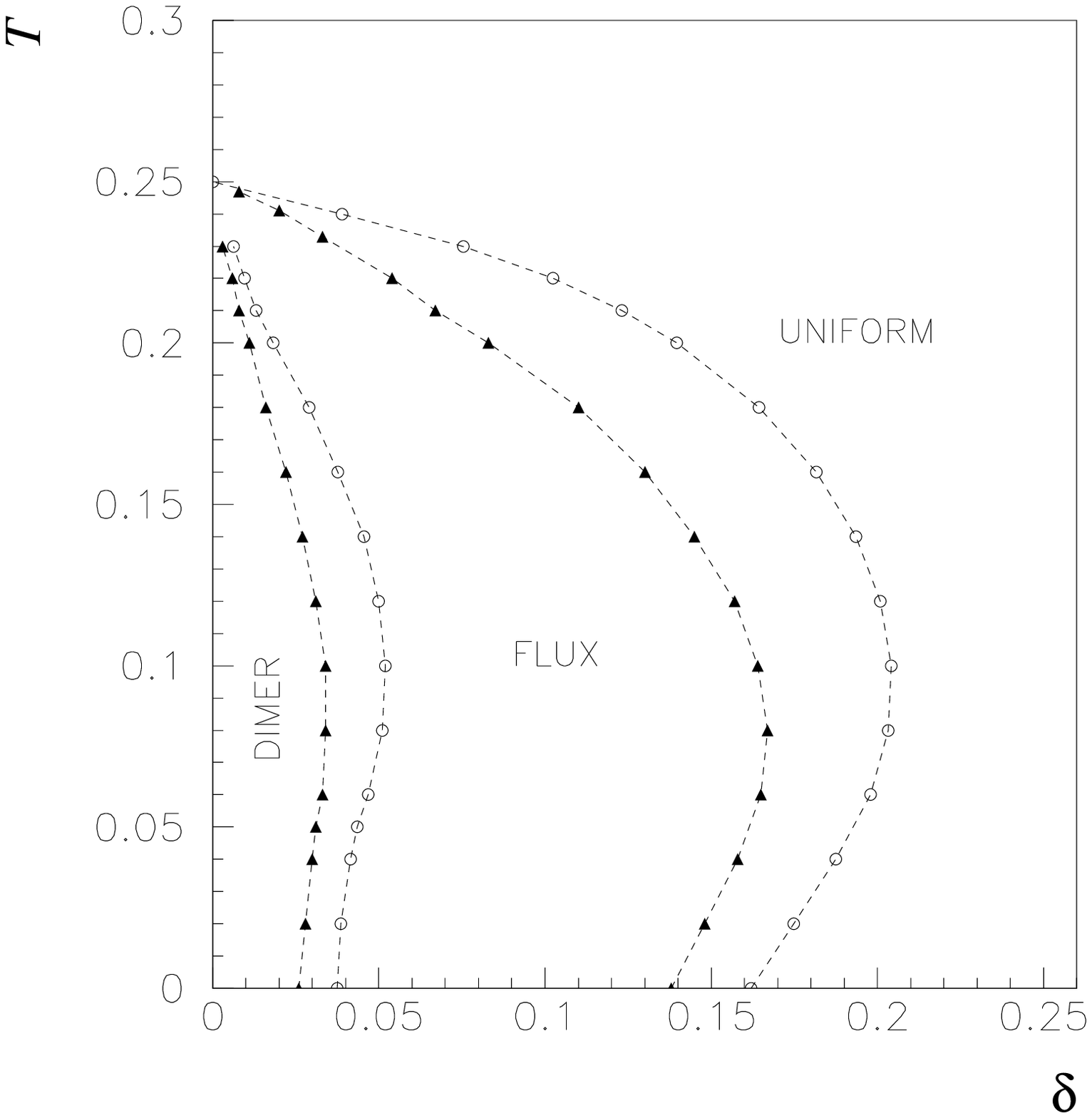}
\caption{\label{fi:1}}
\centerline{{\footnotesize Mean field phase diagram for the $t$-$J$ model 
(open circles)}}
\centerline{{\footnotesize and with the inclusion of the three-site term
(solid triangles).}} 
\end{center}
\end{figure}
\vskip1cm 

Since the hole mobility is enhanced by the addition of the three-site 
hopping term, we expect disordered phases be favored with respect to 
the ordered ones. 
Moreover, the three-site term being proportional to $\d(1-\d)$, 
this effect should increase with doping in the region we considered.
This is in fact what we find in our mean field calculation. 
Fig. 1 shows the shifts of the phase boundaries for the usual $t$-$J$ 
model (open circles) after the inclusion of the three-site term (solid triangles). 
The region of stability of the dimer phase shrinks in favor of flux phase, 
while the uniform configuration extends to considerably lower 
dopings at the expense of the flux one. 

%%%%%%%%%%%%%%%%%%%%    I    %%%%%%%%%%%%%%%%%%%%%%%%%%%%%%%%%%%%%%%%
As we have observed also in \cite{EPR}, for $T=0$ our mean field calculations 
are consistent with the ones carried out within a slave boson formalism
in \cite{UL,GCK}. In particular, we agree with \cite{GCK} in finding 
a first order phase transition between the dimer and the flux phases.
Contrary to \cite{GCK}, we find that the flux-uniform phase transition is 
of the second order for any temperature. This can be seen from either 
Fig. 3 or Fig. 5 of \cite{EPR}.
In Fig. 3 of \cite{EPR} it is possible to observe that, increasing $\d$, 
the angle $\f$, which distinguishes the flux phase from the uniform, 
decreases continuously from $\pi/4$ to 0.
In Fig. 5 of \cite{EPR}, it is easy to see 
that, at the transition point, 
there is no jump in the chemical potential, but only a cusp.
At this regard, we can also observe that the slope of these curves becomes 
steeper as the temperature decreases.
For $T=0$ this behaviour gives rise to numerical problems and probably 
this is the reason why in \cite{GCK} the authors found a jump 
in the chemical potential and hence concluded that the flux-uniform 
boundary line describes a first order phase transition. 
%%%%%%%%%%%%%%%%%%%%%%%%%%%%%%%%%%%%%%%%%%%%%%%%%%%%%%%%%%%%%%%%%%%%%

The diagram of Fig. 1 is constructed without taking into account the 
possibility of coexistence of different phases. Indeed we find 
%%%%%%%%%%%%%%%%%%%%%%
two ranges, belonging respectively to the dimer and the flux phase,   
%%%%%%%%%%%%%%%%%%%%%%
where $\paf{\mu}{\d}>0$ \cite{EPR,PLR} which clearly indicate an instability 
towards phase separation. By means of the Maxwell construction on the chemical 
potential \cite{gri}, we obtain two different regions corresponding to 
dimer-flux and flux-uniform phase coexistence. In the former, the hole-poor 
(hole-rich) part is in the dimer (flux) phase while in the latter the  
hole-poor (hole-rich) part corresponds to the flux (uniform) phase.
The final phase diagram is shown in Fig. 2.

\begin{figure}[htbp]
\begin{center}
\leavevmode
   \epsfxsize=0.35\textwidth\epsfbox{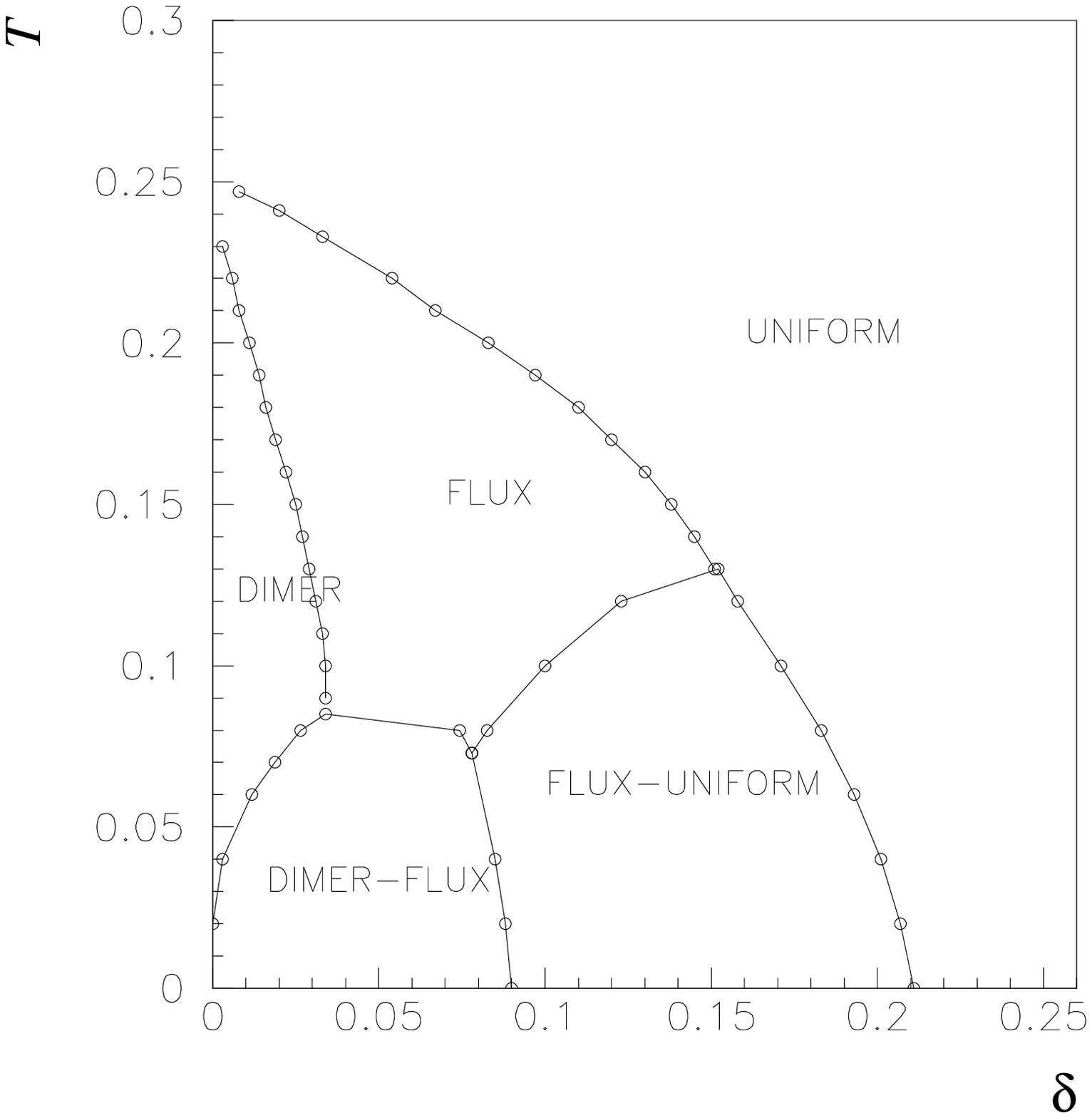}
\caption{\label{fi:2}}
\centerline{{\footnotesize The phase diagram 
after taking into account phase separation.}} 
\end{center}
\end{figure}
\vskip1cm 

%%%%%%%%%%%%%%%%%%%%%%%%%%     II     %%%%%%%%%%%%%%%%%%%%%%%%%%%%%%%
Phase separation for nonmagnetic phases is also 
discussed in \cite{GCK}.  
In the range of dopings 
relevant to high-$T_c$ superconductivity, the authors of \cite{GCK} 
found a wide region of phase separation between a half-filling dimer 
phase and a hole-rich {\it uniform} phase, while we find a smaller 
region of phase separation between a half-filling dimer phase and a 
{\it flux} phase corresponding to $\d\approx 0.09$.  
This discrepancy is due to a different behaviour of the chemical potential 
as a function of $\d$.
Indeed, within the large-$N$ slave-boson formalism 
the electronic chemical potential gets shifted by the mean value of 
the Lagrange multiplier that enforces the constraint, while in the approach 
we adopted such a term is absent. 
Phase separation is a relevant issue because 
%%%%%%%%%%%%%%%%%%%%%%%%%%%%%%%%%%%%%%%%%%%%%%%%%%%%%%%%%%%%%%%%%%%%%
there is a considerable experimental evidence that many high-$T_c$ 
superconductors have a regime in which phase separation between
an AFM hole-poor region and a superconducing hole-rich region occurs 
\cite{HUN}. 
Also, according to \cite{Per} an instability toward phase 
separation might explain the unusual properties of the normal state and 
superconductivity in copper oxides.
%%%%%%%%%%%%%%%%%%%%%%%%%%     II     %%%%%%%%%%%%%%%%%%%%%%%%%%%%%%%
This is why the above mentioned discrepancies between different approaches 
deserve a deeper investigation, which we plan to describe in a future work. 
%%%%%%%%%%%%%%%%%%%%%%%%%%%%%%%%%%%%%%%%%%%%%%%%%%%%%%%%%%%%%%%%%%%%%

\begin{figure}[htbp]
\begin{center}
\leavevmode
   \epsfxsize=0.35\textwidth\epsfbox{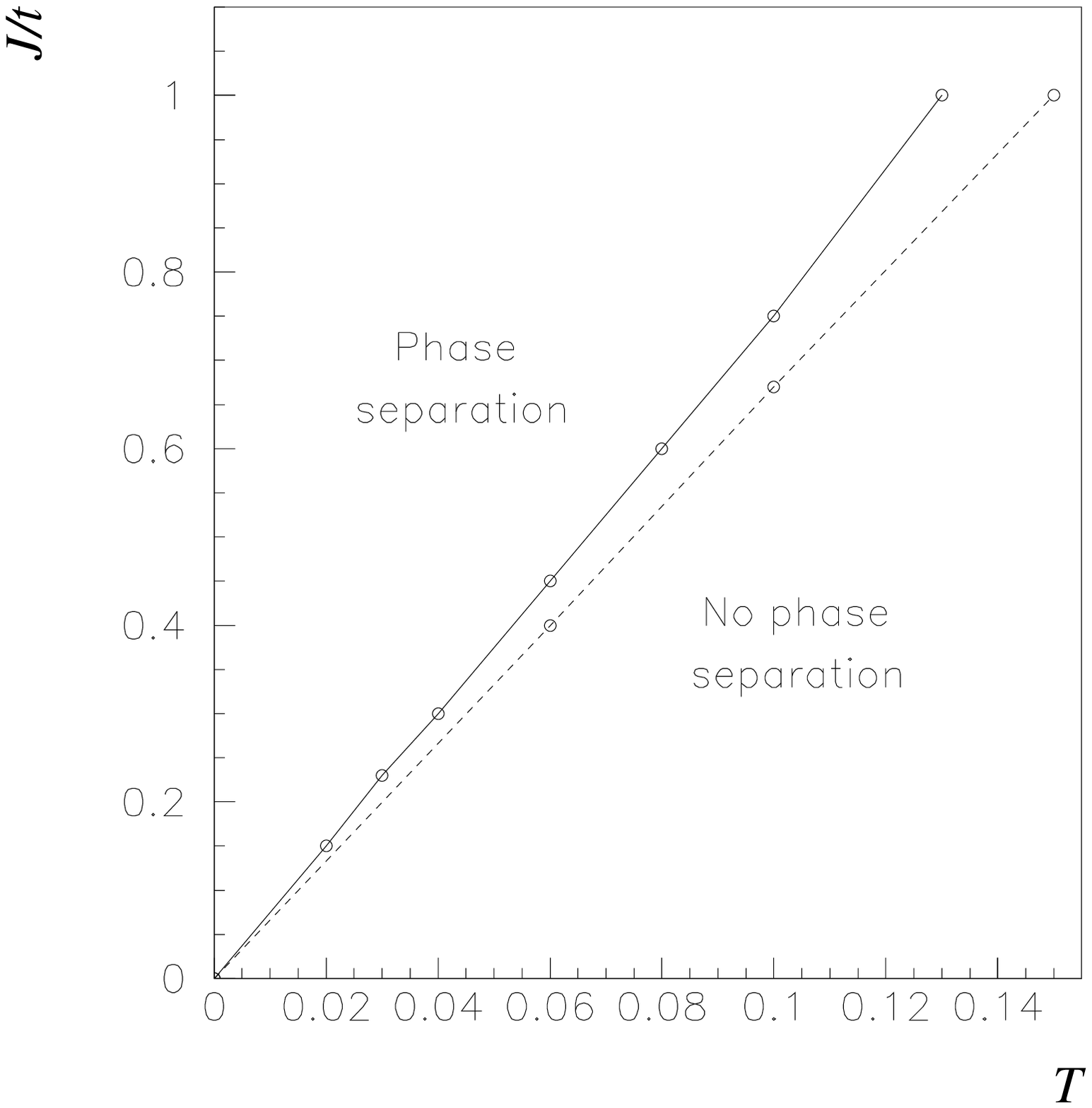}
\caption{\label{fi:3}}
\centerline{{\footnotesize 
The region of antiferromagnetic couplings where phase separation occurs.}}
\centerline{{\footnotesize 
The dotted line refers to a calculation without the three-site term.}} 
\end{center}
\end{figure}
\vskip1cm 

%%%%%%%%%%%%%%%%%%%%%%%%%%%%    III    %%%%%%%%%%%%%%%%%%%%%%%%%%%%%%%
We want finally to give our contribution 
to the open question about the possibility that phase separation 
disappears once $J/t$ is lowered  below a critical value $J_c/t$. 
The determination of this quantity is 
important in order to understand if the $t$-$J$ model can exhibit 
phase separation in the regime of high-$T_c$ superconductors where 
$0.3<J/t<0.5$.
Theoretical studies, for $T=0$, based on different techniques, 
lead to contradictory results. 
Using a high temperature expansion the authors of \cite{PLR} find 
the model phase separates only for $J/t>1$, i.e. outside the region 
interesting for superconductivity. Similar results have been obtained 
in some numerical works \cite{SCL}.
On the contrary other authors 
find that phase separation occurs for all values of $J/t$. 
In particular this has been done by means of quantum Monte Carlo calculations
\cite{HM}, a combination of numerical and analytical results \cite{DR,EKL} and 
the above mentioned slave-boson approach \cite{gri,GCK}.
We calculated numerically the critical value $J_c/t$ for several temperatures 
($T=0$ included) by observing the disappearance of the instability 
regions where $\paf{\mu}{\d}>0$. 

The results, plotted in Fig. 3, show that $J_c/t$ 
varies linearly with the temperature and vanishes for $T=0$, 
both for the standard $t$-$J$ model (dotted line) and with the inclusion of 
the three-site term (solid line). 
Thus, at zero temperature we find phase separation for all values 
of the interaction strength, in accordance with \cite{EKL,GCK,HM}. 
In addition, we can also conclude that at finite temperature there is 
always a nonzero critical value $J_c/t$ below which phase separation disappears.
In particular there is no phase separation in the region interesting for 
superconductivity for temperatures higher than $T\approx0.07t$.

Lastly, we observe that in one dimension some exact diagonalization 
and Monte Carlo results \cite{ATT} suggest  
that the inclusion of the three-site term suppresses phase separation, 
which is instead present when such a term is not added \cite{OLSA, AW}.
This seems not to be the case in two dimensions, at least at mean field
level. Indeed, our calculations show that phase separation 
is still present and the effect of the three-site term is not so dramatic.

%%%%%%%%%%%%%%%%%%%%%%%%%%%%%%%%%%%%%%%%%%%%%%%%%%%%%%%%%%%%%%%%%%%%%%%%%%

%--------------------------------------------------------------
\vskip 0.3cm
{\bf Acknowledgments}

We are indebted to G. Morandi for many valuable suggestions. 
We are also grateful to D. Baeriswyl, F. Napoli and G. C. Strinati  
for helpful discussions. Finally, we would like to thank the referee 
for suggesting us to look for the occurrence of phase separation 
also at small values of $J/t$.  
%--------------------------------------------------------------

\end{document}